\documentstyle[twocolumn,prl,aps]{revtex}

\begin{document}
\draft
\preprint{OUCMT-98-8}
\title{Universal Asymptotic Eigenvalue Distribution of Density Matrices and
 Corner Transfer Matrices in the Thermodynamic Limit}
\author{Kouichi Okunishi, Yasuhiro Hieida and Yasuhiro Akutsu}
\address{Department of Physics, Graduate School of Science, Osaka University,\\
Machikaneyama-cho 1-1, Toyonaka, Osaka 560-0043, Japan.
}
\date{\today}
\maketitle
\begin{abstract}
We study the asymptotic behavior of the eigenvalue distribution of the corner 
transfer matrix (CTM) and the density matrix (DM) in the density-matrix 
renormalization group.  We utilize the relationship ${\rm DM}={\rm CTM}^4$ 
which holds for non-critical systems in the thermodynamic limit. We derive the 
exact and universal asymptotic form of the DM eigenvalue distribution for a 
class of integrable models in the massive regime.  For non-integrable models, 
the universal asymptotic form is also verified by numerical renormalization 
group calculations.

\end{abstract}

The density matrix renormalization group (DMRG) invented by S.R. 
White\cite{DMRG} is one of the most important numerical methods developed 
recently.  Due to the remarkable success, the method has now become one of the 
standard methods for studying one-dimensional (1D) quantum 
systems\cite{Quantum} and two-dimensional (2D) classical statistical 
systems.\cite{Classical}  
In spite of the success, little has been understood as for the foundation of 
the method. Studies clarifying the ``origin'' of efficiency of the method are 
important because they lead to various (including 
higher-dimensional\cite{CTTRG})  extension of the method. One example is the 
work of Ref.\cite{CTMRG}, where a relationship between the density matrix (DM) 
and Baxter's corner transfer matrix (CTM)\cite{Bax1,Bax2,Baxter}is pointed out 
and a new algorithm (CTMRG) is devised.  Another example is the work of 
Ref.\cite{Ost-Rom}, where it is pointed out that the DMRG (at its thermodynamic 
limit) is a variational method using the matrix-product-ansatz (MPA) 
wavefuntion as a trial wavefunction. This leads to a direct variational method 
which does not need the DM\cite{Ost-Rom}, and the product-wavefunction RG 
(PWFRG) which fully utilizes the MPA form of the DMRG-fixed-point 
wavefunction.\cite{PWFRG}

The central object in the DMRG is the DM which is made from the groundstate 
wavefunction (resp. maximal eigenvalue wavefunction) of quantum Hamiltonian 
(resp. transfer matrix) by tracing out information of one half of the system. 
Keeping up to a cut-off-eigenvalue eigenstate of the DM, we have a truncated 
basis set consisting of a finite number (conventionally denoted by ``$m$'') of 
bases, to describe the remaining half of the system.

Since the accuracy of the DMRG is determined by the cut-off eigenvalue, it is 
crucially important to investigate the eigenvalue spectrum $\{\omega_m\}$, in 
particular, its asymptotic ($m\rightarrow\infty$) behavior which has not been 
known precisely.  In this Letter, we present the {\em exact asymptotic form} of 
the DM eigenvalue distribution for a class of (non-critical) integrable models, 
and further, make the first systematic study for non-integrable systems 
employing the CTMRG and the PWFRG by which we can efficiently obtain the 
``fixed-point'' (thermodynamic limit of the system) of the DMRG.

Let us first discuss the integrable cases. In these cases, 1D quantum problems 
are equivalent to 2D classical statistical problems: the Hamiltonian of the 
former can be derived by a log-derivative of the transfer matrix of the latter, 
and the ground-state wavefunction of the former is identical to the 
maximal-eigenvalue eigenfunction $\Psi_{{\rm max}}$ of the latter.\cite{Baxter}  
Hence, we discuss only 2D classical cases in the below.

As has been pointed out by Baxter,\cite{Baxter} the wavefunction (WF) 
$\Psi_{{\rm max}}$ is interpreted as a product of two CTMs, in the 
thermodynamic limit. Since the DM is just a square $\Psi_{{\rm max}}^2$ (with 
$\Psi_{{\rm max}}$ being regarded as a ``wavefuntion matrix''), this 
interpretation leads to a relationship\cite{CTMRG} between the DM, the WF and 
the CTM for 2D classical systems (at least non-critical case where the boundary 
effect is negligible), which is symbolically written as
\begin{eqnarray}
{\rm WF} = ({\rm CTM})^2, \nonumber\\
{\rm DM} = ({\rm CTM})^4. \label{dmctm}
\end{eqnarray}
For integrable models, diagonal form of the CTM is easily known, from which we 
can obtain, for example, the exact one-point function (spontaneous 
magnetization, etc).\cite{Baxter}  Due to the relationship (\ref{dmctm}), the 
diagonal form is also useful to obtain the exact eigenvalue spectrum of the DM.

We discuss the simplest case where the diagonal form of the CTM is given by a 
single infinite tensor product;\cite{Baxter} due to (\ref{dmctm}), diagonal 
form of the DM has the same infinite-tensor-product form with redefined 
parameter. For definiteness let us consider two cases (Type I and Type II) 
where the exact diagonal form of the DM is given by
\begin{equation}
\rho^{\rm (diag)}= \bigotimes_{n=1}^\infty \pmatrix{
1 & 0 \cr
0 & z^{c_n}\cr } \label{exctm}
\end{equation}
with $c_n=n$ for Type I models (e.g, transverse-field Ising chain, 6-vertex 
model, eight-vertex model, XXZ-chain and XYZ-chain) and $c_n=2n-1$ for Type II 
models (e.g., the square-lattice Ising model in the conventional (not 
eight-vertex) representation).\cite{Baxter}  The parameter $z$ ($0<z<1$) 
represents ``degree of non-criticality'' (i.e., $z\rightarrow1$, on approaching 
the critical point), and how it relates to ``physical'' parameters depends on 
the model. Note that the DM (\ref{exctm}) is unnormalized. It is ``normalized'' 
in such a way that its maximal eigenvalue $\omega_0$ is unity; we should divide 
it by ${\rm Tr}\rho^{\rm (diag)}$ for conventional normalization.

Due to the tensor-product structure (\ref{exctm}), each eigenvalue of 
$\rho^{\rm (diag)}$ has the form $z^n$ with $n$ ($\geq 0$) being an integer. 
Further, each eigenvalue $z^n$ may have degeneracy $p(n)$.  To study the 
degeneracy structure of the DM, it is convenient to consider ${\rm Tr} 
\rho^{\rm diag} $:
\begin{equation}
{\rm Tr} \rho^{\rm diag} = \prod_{n=1}^\infty (1+ z^{c_n})
=\sum_{n=0}^\infty  p(n) z^n,  \label{gfctm}
\end{equation}
where the degeneracy $p(n)$ is precisely the coefficient of $z^n$ in the 
infinite series. We should note that, taking the degeneracy into account, the 
number of retained bases $m$ in the DMRG should be
\begin{equation}
m = m(n)=\sum_{k=0}^{n} p(k),\label{mn}
\end{equation}
which means that the cut-off eigenvalue of (unnormalized) DM is $z^{n}$ and 
that we should retain all the degenerate bases belonging to this cut-off 
eigenvalue.

Our problem is to obtain the large-$n$ behavior of $m=m(n)$. For this purpose, 
we should know the asymptotic behavior of $p(n)$. The partition theory of 
integers, which has been played an important role for studies of integrable IRF 
(interaction-round-a-face) models, is helpful again.\cite{Andrews}  By $r(n)$ 
we denote the number of partitions of a positive integer $n$ under a 
restriction ``$r$''  Consider the generating function $f(q)$ associated with 
the restricted partition problem. It has been known\cite{Andrews} that for a 
wide class of partition problems, $f(q)$ can also be given in an 
infinite-product form: 
\begin{equation}
f(q)\equiv \sum_{n=0}^\infty r(n) q^n = 
\prod_{n=0}^\infty (1- q^n)^{-a_n}\label{gfprt}
\end{equation}
where each $a_n$ is a non-negative real number. For the Type I case, we have 
$a_n=1$ for $n$ odd, and $a_{n}=0$ otherwise.

The asymptotics of the generating function of the form (\ref{gfprt}) is 
calculated by the saddle point method; $r(n)$ for $n \gg 1$ is then given by 
Meinardus's theorem (cited in Ref.\cite{Andrews}, page 89):
\begin{equation}
r(n)=A n^{\kappa} \exp( B  n^{\alpha/(1+\alpha)} )+(\mbox{less dominant terms}) \label{meinardus}
\end{equation}
where $\alpha$ is the real part of the pole of the Dirichlet series,
\begin{equation}
D(s)\equiv \sum_{n=1}^{\infty}\frac{a_n}{n^s}, \label{dirich}
\end{equation}
and $\kappa$ is given by
\begin{equation}
\kappa=\frac{D(0)-1-\alpha/2}{1+\alpha}.
\end{equation}
Explicit forms of $A$ and $B$ which we have omitted in the above are also given 
by the Meinardus's theorem. For the Type I models, we have $\alpha=1$ and 
$\kappa=-3/4$:
\begin{equation}
p(n)={\rm const }\; n^{-3/4}\exp( B \sqrt{n} ), \label{pasym}
\end{equation}
where $B=\pi/\sqrt{3}$.\cite{Andrews} For the Type II models, a related theorem 
(Ref.\cite{Andrews}, page 99-100, example 10 and 11) assures the same 
asymptotic form (\ref{pasym}) with $B=\pi/\sqrt{6}$.  It is also possible to 
relate Type II models with the Meinardus's theorem (Chap.1 and 6 in 
ref.\cite{Andrews}). We thus have derived the exact asymptotic form of $p(n)$.

 Using (\ref{mn}) and changing the summation into the integration, we finally 
obtain
\begin{equation}
m  \sim  n^{-1/4}\exp( B\sqrt{n} ), \label{msqr}
\end{equation}
for the Type I and II models. How well the DMRG calculation for the $S=1/2$ XXZ 
chain reproduces the asymptotic behavior (\ref{msqr}) is demonstrated in Fig.1. 
In the actual calculation, we have employed the quantum version of the 
PWFRG,\cite{HOA,OHA,Sato-Akutsu} by which we obtain the fixed-point 
wavefunction of the DMRG efficiently.


  We give a comment on the universality of the asymptotic form (\ref{msqr}) 
among the integrable systems.  In the case where $\{a_{n}\}$ forms a periodic 
series or the model itself admits a direct partition-theoretic 
interpretation,\cite{HardHexagon,ABF} the  $\exp( B\sqrt{n} )$-behavior is 
universal (Ref. \cite{Andrews}, Chapter 6, examples 1-16). The exponent 
$\kappa$ may, however, have possibility of model-dependence (due to $D(0)$), 
modifying the prefactor $n^{-1/4}$ in (\ref{msqr}).

Let us now proceed to non-integrable cases, where the exact diagonal form of 
the CTM or the DM is not known.  The DM eigenvalue is no longer given by 
$z^{{\rm integer}}$ with single parameter $z$, or equivalently, $\log(\mbox{DM 
eigenvalue})$ has not equal-spacing distribution. Both the integer $n$ 
characterizing the DM eigenvalue, and the quantity $p(n)$ which is essential in 
the integrable cases lose meaning.  Our first task is, then, to translate the 
result of integrable cases into the one which has meaning also for 
non-integrable cases.

Writing the $m$-th DM eigenvalue (including degeneracy) as $\omega_{m}$, we 
have $n=\log\omega_{m}/\log z$ in the integrable case.  Substituting $n=\log 
\omega_m /\log z$ into (\ref{msqr}), we have
\begin{equation}
m \sim \left(\frac{\log \omega_m}{\log z}\right)^{-1/4}
\exp\left( B\sqrt{ \frac{\log \omega_m }{\log z} }\right), \label{uve1}
\end{equation}
or equivalently,
\begin{equation}
\log\left[ m \left(\frac{\log\omega_m}{\log z}\right)^{1/4}\right]
=B\sqrt{ \frac{\log \omega_m }{\log z} }. \label{uve}
\end{equation}
From (\ref{uve}), we obtain the leading asymptotic form 
\begin{equation}
\omega_m \sim  \exp[-{\rm const.}\; (\log{m})^2], \label{ued}
\end{equation}
where ${\rm const.}=|\log z|/B^2$ for the integrable cases.  Clearly, 
expressions (\ref{uve1})-(\ref{ued}) do not contain the parameter $n$ which is 
specific to the integrable models.

There arises an intriguing conjecture: the asymptotic forms 
(\ref{uve1})-(\ref{ued}) would also apply to non-integrable systems with $B$ 
and $z$ being suitably redefined.  In the ``neighborhood'' of an integrable 
model with small non-integrable perturbations added, we may well expect this 
conjecture to be true:  In spite of the non-integrable perturbations, the 
``stairway structure'' (or degeneracy) in the DM eigenvalue spectrum still 
remains in somewhat smeared-out way, leaving the ``envelope'' of the 
$\omega_{m}$-$m$ curve essentially unchanged.  As a check of the universality 
for the nearly-integrable cases, we made the CTMRG calculations for two 
systems: the square-lattice Ising model at the critical temperature in finite 
external field and the 3-states Potts model slightly below the critical 
temperature (see Fig.2 and Fig.3).  We see clear agreements between the CTMRG 
calculations and the ``universal asymptotic form''.



As a test of the universality of (\ref{uve1})-(\ref{ued}) for systems {far from  
the integrability}, we take the $S=1$ antiferromagnetic Heisenberg spin chain.  
For calculation of the DM eigenvalue spectrum, we employ the quantum version of 
the PWFRG.\cite{HOA,OHA} The results are given in Fig.4, which support the 
universal asymptotic form.
%


%
We have made similar calculation for the $S=1$ bilinear-biquadratic spin chain 
at $\beta = -0.5$ with the Hamiltonian ${\cal H}=\sum \vec{S}_{i}\cdot 
\vec{S}_{i+1} + \beta \sum(\vec{S}_{i}\cdot \vec{S}_{i+1})^2$, whose result 
(not shown in this Letter) also supports the universality of the asymptotic 
form.

To summarize, we have discussed the asymptotic distribution of the 
density-matrix (DM) eigenvalues for non-critical systems (one-dimensional 
quantum and two-dimensional classical), which controls the accuracy of the 
density-matrix renormalization group. Based on the equivalence between the DM 
and the corner transfer matrix (CTM), we derived the exact asymptotic form of 
the DM eigenvalue distribution for a class of integrable models. The resulting 
expression has been rewritten in a  ``universal'' form which does not contain 
quantities specific to integrable models. Numerical-renormalization-group 
calculations using the CTMRG and the product-wavefunction RG have been 
performed for non-integrable models, which shows that the non-integrable models 
actually have the same asymptotic form of the DM-eigenvalue distribution, in 
strong support of the universality of the asymptotic form.

There remains many important problems left for future studies.  A more 
``physical'' explanation to justify the universal asymptotic form is desired.  
How universal the obtained asymptotic form, itself remains to be a question to 
be answered; there may well be different ``universal classes'' of the DM. In 
fact, the valence-bond-solid (VBS) models\cite{AKLT}, have only 
finite-dimensional DMs which sharply contrast to the ones studied in this 
Letter.  Relation between the DM-eigenvalue distribution and the finite-$m$ 
($m$: number of retained bases) behavior of physical (observable) quantities 
has not been known, although there have been a few works discussing the 
``finite-$m$ scaling''(2D classical\cite{finite-m-cl}, transverse-field XXZ 
chain\cite{finite-m-TXXZ}).  Behavior of the DM for critical system is also an 
important subject of study.\cite{critical,troung}  Our study made in the 
present Letter may be a first step for clarification of these 
problems.\cite{TI}

This work was partially supported by the Grant-in-Aid for Scientific Research 
from Ministry of Education, Science, Sports and Culture (No.09640462), and by 
the ``Research for the Future'' program of the Japan Society for the Promotion 
of Science (JSPS-RFTF97P00201). One of the authors (K. O.) is supported by JSPS 
fellowship for young scientists, and (Y. H.) is partly supported by the 
Sasakawa Scientific Research Grant from The Japan Science Society.

\begin{figure}
\caption{PWFRG calculation of density-matrix eigenvalues $\{\omega_{m}\}$ for 
$S=1/2$ antiferromagnetic XXZ chain and comparison with the exact spectrum. We 
take the exchange coupling constants to be $|J_{x}|=|J_{y}|=1$ and 
$|J_{z}|=\Delta=\cosh(1)$. Number of retained bases in the PWFRG calculation is 
$m=207$.}
\end{figure}

\begin{figure}
\caption{CTMRG calculation ($m=200$) of the density-matrix eigenvalues 
$\{\omega_{m}\}$ for the square-lattice Ising model at a critical temperatur 
$T_{c}$ in a small external field $H$. We have also drawn a line corresponding 
to the universal asymptotic form.}
\end{figure}

\begin{figure}
\caption{CTMRG calculation ($m=242$) of the density-matrix eigenvalues 
$\{\omega_{m}\}$ for the 3-state Potts model slightly below the critical 
temperature. We have also drawn a line corresponding to the universal 
asymptotic form.}
\end{figure}

\begin{figure}
\caption{PWFRG calculation ($m=700$) of density-matrix eigenvalues 
$\{\omega_{m}\}$ for $S=1$ antiferromagnetic Heisenberg chain. We have also 
drawn a line corresponding to the universal asymptotic form.}
\end{figure}

\end{document}